# Object Tracking and Identification by Quantum Radar

Kadir Durak*[a,b], Naser Jam[a], Çağrı Dindar[a]
[a]Dept. of Electrical and Electronics Engineering, Özyeğin University, Istanbul, Turkey, 34794;
[b]BİLGEM, Scientific and Technological Research Council of Turkey, Gebze, Turkey 41470

## ABSTRACT

Quantum Radar is a promising technology that could have a strong impact on the civilian and military realms. In this study we introduce a new concept design for implementing a Quantum Radar, based on the time and polarization correlations of the entangled photons for detection and identification and tracking of high-speed targets. The design is focused on extracting high resolution details of the target with precision timing of entangled photons that provides important operational capabilities like distinguishing a target from a decoy. The quantum entanglement properties guarantee the legitimacy of the photons captured by the search telescope. Time correlations of the photon detection events can be extracted via cross-correlation operation between two sets of photon detection time-tags for the entangled photons. The fact that the wavelengths of the entangled photons can be tuned also makes the Quantum Radar concept an enticing candidate for tracking stealth objects. We present the proof-of-principle test results of the Quantum Radar and discuss the technical challenges and limitations of the design.

## 1. INTRODUCTION

In 1955, John S. Bell published his famous thought experiment for testing local realism against Copenhagen interpretation. After that a lot of sophisticated experimental tests have suggested that the local realism assumption is wrong. Nowadays the non-locality is used as a useful resource for novel communication schemes [1-3] and it seems that it can also effectively be used in sensor systems that needs information from a far object. Quantum Radar is a promising technology that could have a strong impact on both the civilian and military applications because stealth and decoy technologies are getting more mature and practical [4]. The traditional Radar signal is made of huge number of photons such that the particle properties of light cannot be observed but it behaves as classical light, i.e. electromagnetic wave. The spectrum is in the radio frequency (RF) range. Quantum Radar generalizes the concept of Radar, but this novel device operates with a relatively small number of photons. As such, its theoretical description must be done using quantum electrodynamics. Quantum Radars may be able to detect, stealth platforms and decoys. In the electronic battlefield, Quantum Radar may become a revolutionary technology just as stealth technology was in the last three decades of the 20th century.

Quantum sensing technologies are getting more practical; however, only a few reports and news are published about Quantum Radars and there was only one claim about a working prototype with 100 km range by China Electronics Technology Group Corporation (CETC) in the Zhuhai Airshow on Nov. 6, 2018 [5], which is essentially an optical Radar (Lidar) claimed to use quantum technology to enhance Radar performance to detect weak photons reflected from the stealth targets. According to a report on 2009 by Air force Research Laboratory [3] and a review paper on 2015 by *Kyriakos Ioannou* [7], other designs are just in proposal or laboratory prototype level and there is no report yet even about a working prototype. After a careful review of published books, papers and news [8-14], we found that previous works on Quantum Sensors are concentrated on using unentangled or entangled photons for enhancement of echo signals received from the targets. According to the type of quantum phenomena exploited by the systems, DARPA classified Quantum Sensor architectures in three basic categories as following [6]:

**Type 1:** The quantum sensor transmits unentangled quantum states of light.
**Type 2:** The quantum sensor transmits classical states of light but uses quantum photo-sensors to boost its performance.
**Type 3:** The quantum sensor transmits quantum signal states of light that are entangled with quantum ancilla states of light kept at the transmitter.

*kadirac@gmail.com;     phone: +90 216 564 93 67;     fax: +90 216 564 99 99;     www.ozyegin.edu.tr

To date it is very difficult to predict the performance of a Quantum Radar in an operational environment. Firstly, most Quantum Radar proposals presume a perfectly reflecting target oriented in the specular direction. Another challenge is that we cannot use the standard definition of the radar cross section [8]. In this report, we present proof-of-principle experiment that is suitable for object identification and tracking using entangled photons, i.e. a Quantum Radar.

## 2. CONCEPT AND DESIGN CONCERNS

As our approach does not imply any limitation to the wavelength, it can work either in Radio-wavelengths as Radar; or Laser-wavelengths as Ladar/Lidar. Although, the single photon detection technology is not developed for some bands of light spectrum, theoretically there is no fundamental limitation on the wavelength for the design. So, for the sake of abstraction we use more generalized term "Radar" for this design. Here we show that the enhanced target detection advantage based on the quantum illumination is not the prominent advantage. In most of the previous works the motivation for developing the concept of Quantum Radar was the potential use of quantum phenomena to increase the sensitivity of the sensing system, because, entangled-photon Quantum Radar appears to offer higher sensitivity.

Figure 1 shows the block diagram of designed Quantum Radar that uses both entangled and untangled single photons to increase signal-to-noise ratio and time resolution in the same time. Unlike previous Quantum Radar proposals our design does not focus on the sensitivity only and uses coincidence detection[15] and entanglement property; based on statistical signal processing on timing of reflected single photons and violation of Bell's inequality, respectively, to distinguish between a real target and a decoy. This requires Radar's time measurements resolution extremely high in a way that its limit is our time-stamp hardware's resolution and relativistic effects of the fast-moving target on time-of-flight. For instance, if resolution of time-stamp hardware is 100 picoseconds, the Radar's system-level range resolution should be in the centimeter range. In this approach, targets are discriminated by using successive cross correlation of received photon timings with multiple Doppler shifted and relativistic time delated replicas to evaluate target's range, speed and form.

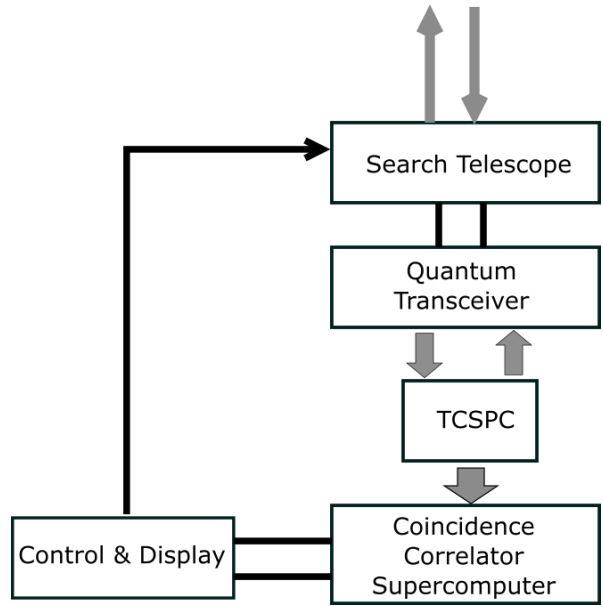

Figure 1. Block diagram of quantum radar design.

### 2.1 Quantum Radar Concept

If we assume that we need to illuminate the entire body of a typical target with 10 m span in 10 km Range, then the radiant angle will be: $\Theta_B = Span / Range = 10/10^4 = 1MIL$. As shown in Figure 2 the Radar repeatedly sends single photons to space by a telescope and evaluates the received photons in the following way:

1. One of the photon pairs, say signals, are kept as reference and their compressive timing table for K=1 to N are recorded,
2. Transmitter sends N idler photons and makes a combination of delta functions $D_T(t)$ to represent it as a time signal

$$D_T(t) = \sum_i \delta(t-t_i)$$

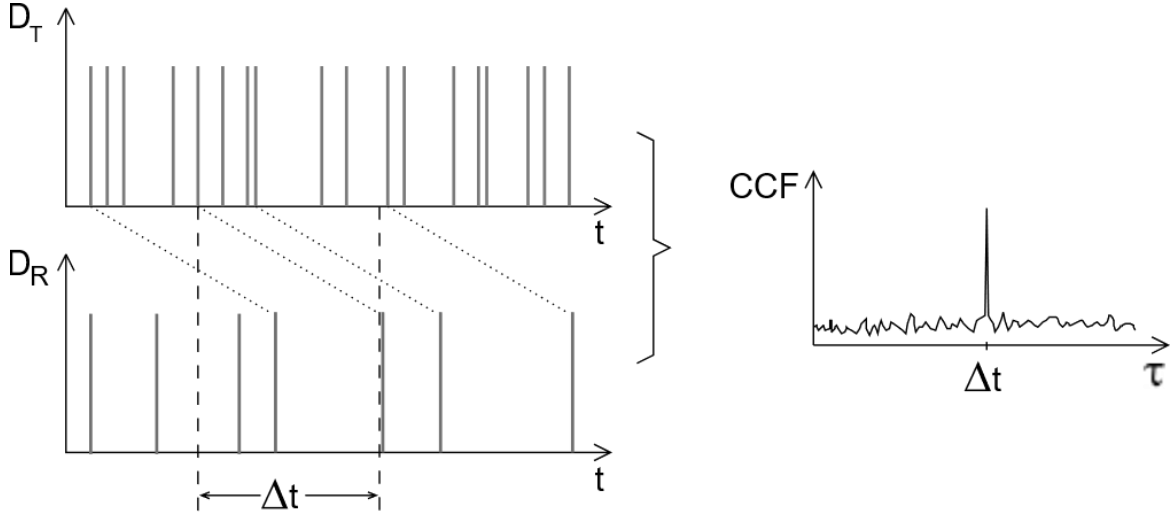

Figure 2. Extracting the target range by cross correlation of timing signals.

3. When the idler photons are back-scattered from the object, they are captured by the telescope and the receiver digitizes the output signal of the detectors as $D_R(t)$
4. Receiver calculates the cross-correlation function $ccf(\tau, \gamma, f_d)$ for all possible time delays, doppler shifts and all possible relativistic time dilations.

$$ccf(\tau, \gamma, f_d) = K\int^T D_T(t)\, D_R(t+\tau, \gamma, f_d)dt$$

in which
$ccf(\tau, \gamma, f_d)$ means the cross correlation of the transmitted pulses with received signal plus noise,
$\tau$ is the time shift (due to optical delay) in correlation,
$f_d$ is the Doppler shift of the received signal and
$\gamma$ is the relativistic time dilation of the reflected photons known as Lorentz factor and is given by:

$$\gamma = \frac{1}{\sqrt{1-\beta^2}}$$

Here $\beta = v/c_a$, which is relativistic effect's approximation at low speeds that holds to within 0.1% error for $v < 0.22\, c_a$ ($v < 66{,}000$ km/s). (where v is the speed of the target and $c_a$ is speed of photon in the air.) For instance, for a hypersonic target speed of 8 Mach (2722 m/s) the time difference should be 0.0412 ns. Which is equal to 6 mm resolution for extraction of target geometry.

5. The receiver finds the peak of the cross-correlation function revealing the precise round-trip transit time of the single photon $t_r$ that is related to the distance of the target R by: $R = c_a\, t_r\, /2$, which means the distance to the target is half the round-trip transit time multiplied by the speed of light in the air. In this way we can find the location of the target in polar coordinates by evaluating the azimuth and elevation angles of the telescope ($\Theta$, $\phi$). Velocity of the target is also extracted by $f_d$ (doppler frequency bin in which cross-correlation peak occurs) that is directly related to relative velocity $v$ of the target by: $v = c_a\, f_d\, /2PRF$, in which PRF is the transmitted Photon Repetition Frequency. Figure 3 visualizes typical results of the coincidence correlation for different doppler shift bins.

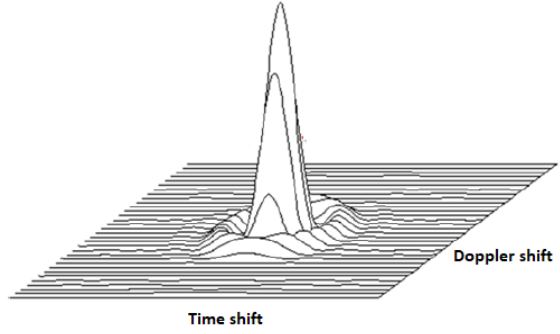

Figure 3. Coincidence correlation in time domain and frequency domain

## 2.2 Relativistic Concerns

For a source of light at a given frequency, at rest; it emits every $t = 1/v$ seconds a new wavefront. $\lambda$ is the distance between two wavefronts. An object moves towards the source with the speed of u. If we would like to put Special Relativity into the consideration, then the time for the object would be $t = \dfrac{t_0}{\sqrt{1-\beta^2}}$ relative to the time for observer $t_o$.

Then, the shift would actually be $t_{cls}' = \dfrac{1-\beta}{t}$ where $\gamma = \dfrac{1}{\sqrt{1-\beta^2}}$. The actual shift on the frequency would be [16]

$v' = v\sqrt{\dfrac{1+\beta}{1-\beta}}$. Now, let's consider that for each wavefront, we have a single photon. Then, we can say that the source produces a new photon every $t = 1/v$ seconds. Then, the time shift because of classical Doppler shift from would be $t_{cls}' = \dfrac{1-\beta}{t}$ and the time shift because of relativistic Doppler shift would be $t_{rel}' = t\sqrt{\dfrac{1-\beta}{1+\beta}}$. So, the change at the doppler time shift $\Delta t$ when the special relativity taken into account is: $\Delta t = \Delta t_{classical} - \Delta t_{relativistic}$, where $\Delta t_{classical} = t - t_{cls}'$ and $\Delta t_{relativistic} = t - t_{rel}'$. It is then concluded that the special relativistic corrections should be done for the cross correlation of timing signals. Practically for the sake of efficiency, the above-mentioned detection process is performed in frequency domain by taking the Fast Fourier Transform of received signal and multiplying it with doppler shifted frequency-domain replicas of transmitted signal $D_T(t)$.

$$ccf(\tau, \gamma, f_d) = F^{-1}\,[F[D_T(t)]\, F\,[D_R(t+\tau, \gamma, f_d)]]$$

## 2.3 Target identification with cross-correlogram

The identification process of the system is based on cross-correlogram processing. This kind of identification process simplifies the optical sections and eliminates the need for complex micro-scan devices, that make the system very expensive, complex and dependent on the precise mechanical tolerances. In fact, instead of scanning fine angles in the space to extract details of the object we use quantum mechanical nature of the single photons to scan tiny angles of space in a probabilistic way.

The concept of the process is shown in Figure 5. The photons are randomly reflected from different parts of the target's surface with different time of flight, then the system makes a cross-correlogram of photons in each spherical angle. The expansion of this correlogram will show the depth of the target. A decoy which has a shorter length, its correlogram should have less span in time-of-flight domain, but a real target which has a larger body its correlogram should be wider. Accordingly, a neural network that is well trained with the several samples of 3D CAD drawings of targets and typical decoys can provide a reliable threshold for differentiating between targets from decoys.

## 2.4 Analogy and Comparison to Classical Radar

Quantum Radar's detection is based on input quantum correlations and output quantum detections. As discussed before one of the advantages of Quantum Radar is claimed to be detection of stealth targets. Because if it is carefully engineered, it will extract its transmitted signal from the background noise thanks to entanglement properties, i.e. polarization and time correlations. This allows it to detect stealth aircraft, filter out wideband jamming and operate in the areas of high background noise. Another important advantage is that, single photons are not easy to detect, which gives no warning to targets to be detected. This means that, unlike a classical Radar, which exposes itself to detection whenever it actively emits radio waves to search for aircraft, the Quantum Radar does not announce its location to anti-radiation missiles. By exploiting entanglement, Quantum Radar offers the prospect of enhanced target detection capabilities. These systems rely on quantum states of photons sustained on an entangled superposition. An entangled state is generated in the Radar. The signal (idler) photons stay in a controlled transmission line while the idler (signal) photons are emitted towards the possible object we want to detect. Since its reflectivity is small, $\sigma \ll 1$, most of the light captured by the receiver is background noise. By measuring the correlations between the signal and the idler photons, it is possible to detect the presence of an object with a smaller error probability than protocols involving classical light, with a gain up to 3 dB in the error probability exponent. This is based on this fact that Entanglement can enhance the distinguishability of entanglement-breaking channels [17]. Entanglement-based Quantum Radars have their limitations; like traditional Radars, they degrade in resolution over longer distances as well as the performance dependency on the weather conditions. This is because the entangled particles do eventually lose the coherence of their quantum state over long distances due to atmospheric turbulences and other decoherence agents, a phenomenon which can worsen in adverse weather.

## 2.5 Jam and Decoy Resistance of Quantum Radar

It can be shown that the most important advantage of this Quantum Radar over classical one is the electronic warfare superiority; it is quite like the same advantage that we have in Quantum communications. In this case if we use single photons instead of classical electromagnetic wave for target detection and tracking, then it is impossible to imitate our Radar signals and the Radar could not be deceived by fake signals. However, active radar countermeasures include the use of jamming to hide the presence of a target. In this regard, it is important to mention that Quantum Radar appears to be more robust against the effects of a jamming signal. Indeed, let us recall that both Quantum Radar proposals, the interferometric Quantum Radar and quantum illumination, take advantage of the same principle to achieve higher performance. These systems perform collective measurements after highly correlated states have been injected into the system. These correlations can be exploited to prevent signal photons to get lost in noise or jamming signals. Let's consider, the case of quantum illumination using non-entangled photons.
The signal-to-jam ratio will be approximately given by:

$$S/J \approx K \sigma / P_j$$

where $\sigma$ is the cross-section of the target and $P_j$ is the average number of photons transmitted by the jammer.
On the other hand, if we use entangled photons, the signal-to-jam ratio is now given by:

$$S/J \approx K \ 2^m \sigma_Q / P_j$$

where $\sigma_Q$ is the quantum cross-section and *m* is the number of qubits of entanglement in the photon states [5]. Consequently, entangled quantum illumination is less likely to be jammed than a Quantum Radar operating with non-entangled photons. That is, in terms of entanglement as a resource, a Quantum Radar is exponentially less likely to be jammed than a classical Radar. The most effective advantage of this design is its capability to distinguish between decoys and real targets. This design might be capable of differentiating between the materials of targets due to the way single photons interact with surfaces they contact. This could allow more detailed identification; helping distinguish between decoys and real targets. This could be a serious concern as there is a market on low-cost Air Launched Decoys (shown in Figure 4) that can easily dismantle advanced surface-to-air missile systems like S-400 or Patriot Air defiance System that can resist in an electronic-warfare-heavy environment. For example, Raytheon's Miniature Air Launched Decoy (MALD), a sleek, 280-pound cruise missile can mimic various aircraft with radar reflectors. The MALD-J version has electronic jammers, and the combination of decoys and jammers can "deceive, distract and saturate" radar systems with false signals, the company says. Drone decoys can get close and operate within the "no escape range" of radar systems, something too dangerous for manned aircraft [18].

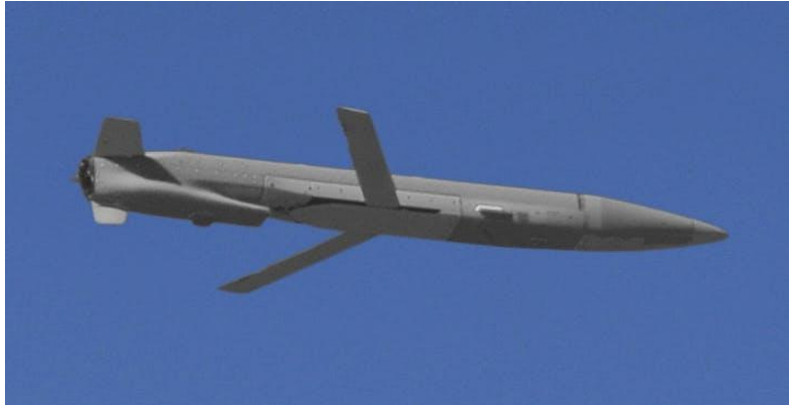

Figure 4. Air launched decoy.

## 3. EXPERIMENT AND RESULTS

For evaluation of the latest state-of-the-art technologies and their impact on precision of the system, we are using an experimental setup to implement a short-range primitive of the Quantum Radar. In this setup, we use an entangled photon source for illumination of the test object. While one arm of the pairs are measured by single photon counter and their time-stamp information is kept as reference, the others are sent into free-space for a possible object detection. Only a small portion of the reflected photons from the object are received by the telescope and they are also detected. The received photons' timing is captured by the event timer of the time-stamp unit. This part of device is in fact a timing to digital converter with time resolution of 81 picoseconds. The acquired timing values are fed to the correlation engine by a fast USB3 port in which coincidence correlation is calculated for fixed objects and moving objects. Figure 5 shows the experimental setup with the object to be detected. Laser diode (LD) creates continuous wave (CW) light with a central wavelength of 405 nm and linewidth of 160 MHz. Flourescence filter (FF) is used to filter unwanted stray fluorescence and half-wave-plate ($\lambda/2$) prepares the pump beam's linear polarization. Nonlinear crystal (NC) is Beta Barium Borate (BBO) crystal with a cut angle of $28.78^0$. Down conversion event allows the creation of signal and idler photons of central wavelength values 780 nm and 842 nm, respectively. Low pass filter (LF) filters the blue light from the system and only down converted photons are left in the optical path. Dichroic mirror separates the signal and idler photons and sends the signals to counter (C1) for time-tagging as reference. Polarizing beam splitter (PBS) transmits the horizontal idler photons and a quarter-wave-plate ($\lambda/4$) is used to change the polarization to right-hand circularly (RHC) polarized light. Then the telescope sends the RHC polarized idler photons to object to be detected. During the back-scattering the handedness of the idler photons changes and it becomes left hand circularly (LHC) polarized. Now the $\lambda/4$ converts the idler photons to vertically polarized ones, which gets them reflected on the PBS and detected at C2. The time-stamp information of idler photons is also recorded to make an analysis with the reference to identify the coincidences. The time delay that gives the maximum coincidence value gives the distance between the telescope and the object. If the object moves with a certain speed, the time delay between signal and idler time-stamps would require a uniformly varying delay instead of a constant one. In this proof-of-principle experiment we only tested against stationary objects. The dynamic object detection and tracking requires rather powerful computers.

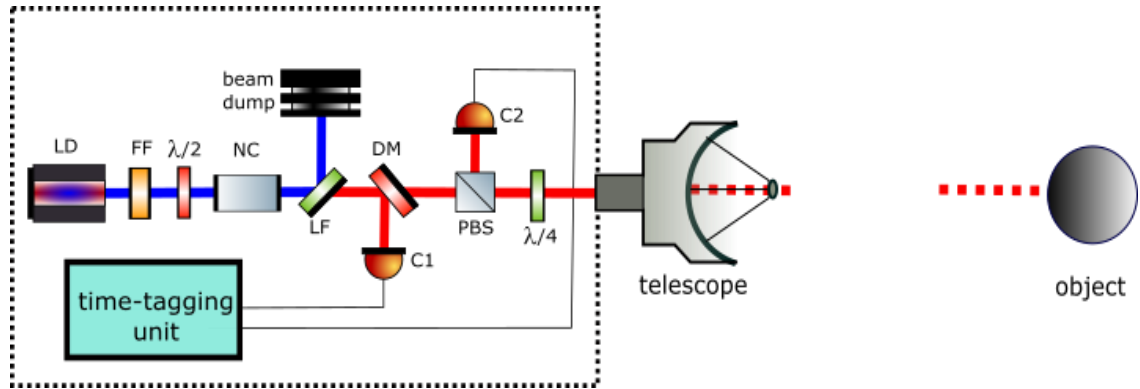

Figure 5. Schematic of the experimental setup to test the proof-of-principle quantum radar concept.

We have chosen black anodized Aluminum as the object material due to its good absorptivity which is near unity in the visible/near infrared (NI) spectrum [19]. The absorptivity and scattering are measured to be 0.87 ± 0.02 and 0.13 ± 0.03, respectively. Although, the exact scattering nature of the object depends on the surface roughness of the material, the amount of detected back-scattered light is expected to decrease with the inverse square of the distance between the telescope and the object, and quadratically increase with the aperture of the telescope: $P \sim D^2/R^{-2}$, where P is the back-scattered power, D is the telescope diameter and R is the distance between the telescope and the object. This suggest that: i. the signal reduces significantly with increased distance to object and ii. the telescope diameter can be increased to cover larger distances with the Quantum Radar. The largest known diameter values of optical telescopes are few meters scale. However, higher entangled photon creation rate and detection rate values can help increasing the range of the Quantum Radar as well. Fortunately, both can significantly be increased with various multiplexing techniques. For the rate matters the limitation comes from the timing jitters of the single photon detectors. Higher photon rate values require smaller coincidence window so that more coincidence events can be measured. A coincidence is the occurrence of two photon detection events to be within a time window. However, the timing jitter of commercially available avalanche photodiodes (APD) is typically on the order of hundreds of picoseconds. If the coincidence window is chosen to be smaller than the timing jitter, the coincidence will be missed. Therefore, with the typical APDs the limit would be on the order of $10^{10}$ photon pairs per second. The suppression of the timing jitter for APDs is an area of active research and there are ongoing efforts to reduce it to tens of ps [20].

In our experimental setup, we can detect 5 millions of photon pairs with our single photon counters (SPCs), which have photon detection efficiencies of 53% ± 1.2 and 55% ± 0.65. Therefore, the coincidence to single (or pairs to single) ratio with these detectors would be around 0.25. We measure the scattered photons from black anodized Aluminum post holder of 25 mm diameter, 50 mm diameter and a flat base plate. The objects are chosen according to the ease of access in the laboratory environment, not for a specific experimental purpose. The coincidence rates are measured by finding the correct time delay between entangled photons by cross-correlation operation. The fit follows the equation of the form $f(x)=b+ax^{-2}$, where b represents the background (or accidental) coincidence rate and a is the coefficient that accounts for all the attenuations of the signal, x is the distance in mm from the object to the SPC and the function shows the coincidence rate (fit parameters are: a=75.14 and b=78). The fit parameters highly depends on the telescope aperture, which is 50 mm in our experimental setup.

The fit indicates that if we had 10 billion of entangled photon pairs per second above 300 meters of distances, the coincidence rate goes down to single digits. Therefore, the experimental data and its extrapolation indicates that with the existing APD technologies the maximum achievable range for a Quantum Radar would be below 1 km if the target to be detected is black anodized aluminum. However, different materials with better scattering and lower absorption may allow higher range target detection. The diameter of the telescope extends the working range quadratically.

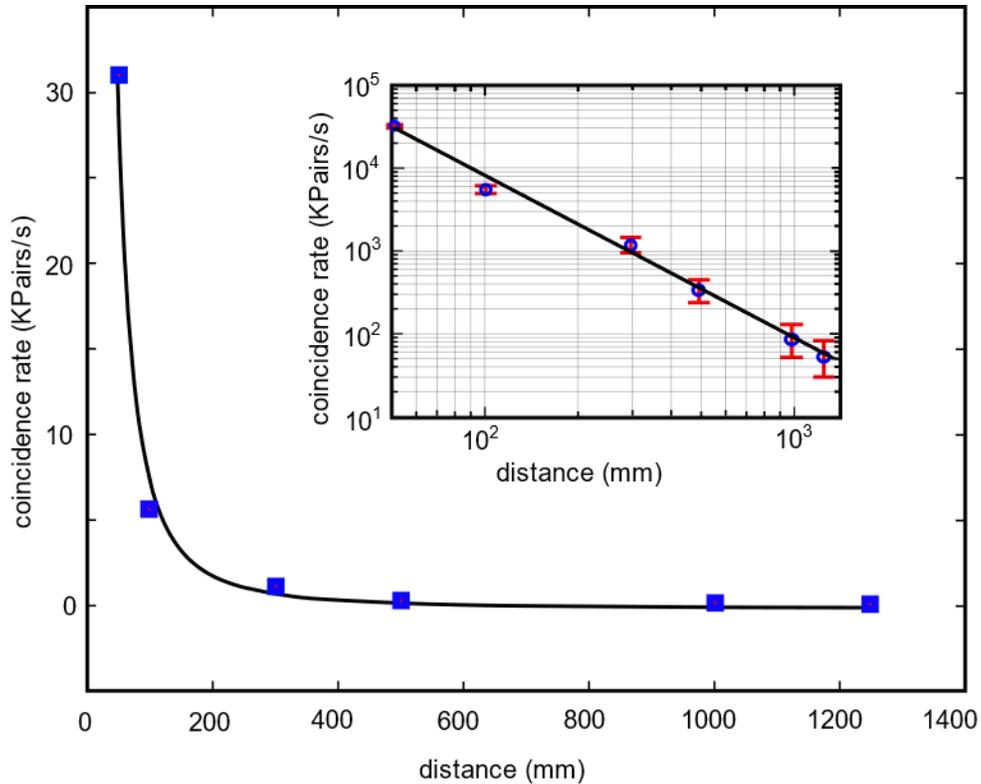

Figure 6. Cross-correlogram delay values are converted to distance. The solid black line shows the theoretical fit to the experimental data. The inset shows the logarithmic scale for the ease of reading.

## 4. CONCLUSION

In this design and experiment we have explained the electronic warfare superiority of a specially designed Quantum Radar for distinction between targets and decoys. We showed that instead of using less sensitive and complex 3D Raster scan single photon Lidars, by using this approach we can use the statistical nature of photon scattering from a target in a relatively large field-of-view to extract precise information of the target geometry required for distinctive correlations. We have also showed that for the sake of precision how we can take into the account the relativistic time dilations which effect on the time-of-flight of photons for high speed supersonic and ballistic targets. Although it has superior performance on being undetectable and identifying decoys, the current electro-optic technology limits the Quantum Radar design range to few hundreds of meters with centimeter scale telescope aperture. However, it should be noted that the limitation is not due to fundamental physics but only a result of current challenges of opto-electronic technology. There are ongoing efforts on advancing the APD technologies which can possibly extend the current limits of Quantum Radar. With such advances in APD technology and large telescope aperture the presented design will be a promising technology.